%% file: arxiv.tex
\documentclass[
reprint,
superscriptaddress,
amsmath,
amssymb,
pra
]{revtex4-2}

\usepackage{graphicx}
\graphicspath{{../figures/}} 
\usepackage{dcolumn}
\usepackage{bm}
\usepackage{physics}
\usepackage{braket}
\usepackage{placeins}
\usepackage[colorlinks=true]{hyperref}
\usepackage{upgreek}
\usepackage{multirow}
\begin{document}

\let\oldaddcontentsline\addcontentsline
\renewcommand{\addcontentsline}[3]{}

\input{main_header.tex}

\date{\today}

\begin{abstract}
   We use electronic microwave control methods to implement addressed single-qubit gates with high speed and fidelity, for $^{43}$Ca$^{+}$ hyperfine ``atomic clock'' qubits in a cryogenic (100K) surface trap. For a single qubit, we benchmark an error of $1.5$ $\times$ $10^{-6}$ per Clifford gate (implemented using $600~$ns $\pi/2$-pulses). For two qubits in the same trap zone (ion separation $5~\upmu$m), we use a spatial microwave field gradient, combined with an efficient 4-pulse scheme, to implement independent addressed gates. Parallel randomized benchmarking on both qubits yields an average error $3.4$ $\times$ $10^{-5}$ per addressed $\pi/2$-gate. The scheme scales theoretically to larger numbers of qubits in a single register.
\end{abstract}


\maketitle

\input{main.tex}
\bibliography{library.bib}

\FloatBarrier
\clearpage
\onecolumngrid
\begin{center}
{\Large \textbf{Supplementary information}}
\end{center}
\makeatletter
   \renewcommand\l@section{\@dottedtocline{2}{1.5em}{2em}}
   \renewcommand\l@subsection{\@dottedtocline{2}{3.5em}{2em}}
   \renewcommand\l@subsubsection{\@dottedtocline{2}{5.5em}{2em}}
\makeatother
\let\addcontentsline\oldaddcontentsline

\renewcommand{\thesection}{\arabic{section}}

\onecolumngrid

\let\oldaddcontentsline\addcontentsline
\renewcommand{\addcontentsline}[3]{}
\let\addcontentsline\oldaddcontentsline
\renewcommand{\theequation}{S\arabic{equation}}
\renewcommand{\thefigure}{S\arabic{figure}}
\renewcommand{\thetable}{S\arabic{table}}
\renewcommand{\thesection}{S\arabic{section}}
\setcounter{figure}{0}
\setcounter{equation}{0}
\setcounter{section}{0}

\newcolumntype{C}[1]{>{\centering\arraybackslash}p{#1}}
\newcolumntype{L}[1]{>{\raggedright\arraybackslash}p{#1}}

\FloatBarrier

\input{SI.tex}

\end{document}

%% file: main_header.tex
\title{Fast, high-fidelity addressed single-qubit gates\\ using efficient composite pulse sequences}

\author{A.\,D.\,Leu}
\email{aaron.leu@physics.ox.ac.uk}

\author{M.\,F.\,Gely}

\author{M.\,A.\,Weber}

\author{M.\,C.\,Smith}

\author{D.\,P.\,Nadlinger}

\author{D.\,M.\,Lucas}

\affiliation{Clarendon Laboratory, Department of Physics, University of Oxford, Parks Road, Oxford OX1 3PU, U.K.}

%% file: main.tex
%

Trapped ions are one of the most promising platforms to build a universal quantum computer~\cite{Monroe_2013}.
Quantum state control of ions is conventionally achieved with lasers, but radio-frequency~\cite{Srinivas_2021} or microwave fields~\cite{Entangling,Ospelkaus_2011,Weidt_2016,Harty_2014,harty2016,zarantonello2019} have in recent years demonstrated competitive performance.
Microwave technology is more mature and widespread than laser technology and hence cheaper and more reliable.
Also, the long wavelength of microwaves eases phase control, and waveguides can straightforwardly be integrated into surface ``chip'' traps.
Microwave-driven logic is therefore a compelling candidate for scaling up ion trap quantum processors, and gates surpassing error correction thresholds have been demonstrated~\cite{Harty_2014,harty2016,zarantonello2019}.
However, whilst laser beams can be focused to address individual ions in the same trap potential~\cite{laserfocus}, the centimetre-scale wavelength of microwaves requires a different approach to single-ion addressing.
Past demonstrations of microwave-driven addressed gates have mostly relied on nulling the effect of the microwave field for the non-addressed ion.
This has been achieved through position-dependent Zeeman shifts~\cite{Piltz_2014,Warring_2013,Randell_2015} or by nulling the field amplitude for certain ion positions~\cite{Craik_2017,Warring_2013}.
Similarly, sidebands of the microwave qubit transition can be generated and nulled at different positions, either by controlling micromotion~\cite{Warring_2013}, trapping ions in different potential wells with different secular frequencies~\cite{sutherland2022} or by stimulating ion motion using d.c. electric fields~\cite{srinivas2022}.
However, gate errors and crosstalk below $10^{-4}$ -- an important threshold for the practical scalability of error-correction~\cite{knill2010,Preskill1998} -- have not previously been demonstrated.
In this Letter, we first report on global single-qubit operations comparable to the present state of the art~\cite{Harty_2014} but featuring a considerable $\sim20 \times$ speedup.
We exploit this performance improvement to implement a more complex multi-pulse scheme which can address two ions within the same potential well, with an average addressed gate error of $3.4 (3)\times 10^{-5}$ including crosstalk errors and with faster gate speed to that of~\cite{Harty_2014}.
The scheme employs the microwave field gradient in our trap and uses an efficient composite sequence of single-qubit rotations to perform an arbitrary combination of addressed gates on both ions simultaneously.
We characterize the addressing scheme by carrying out independent randomized benchmarking (RB) sequences on both ions simultaneously.

\begin{figure}
    \centering
    \includegraphics[width=0.43\textwidth]{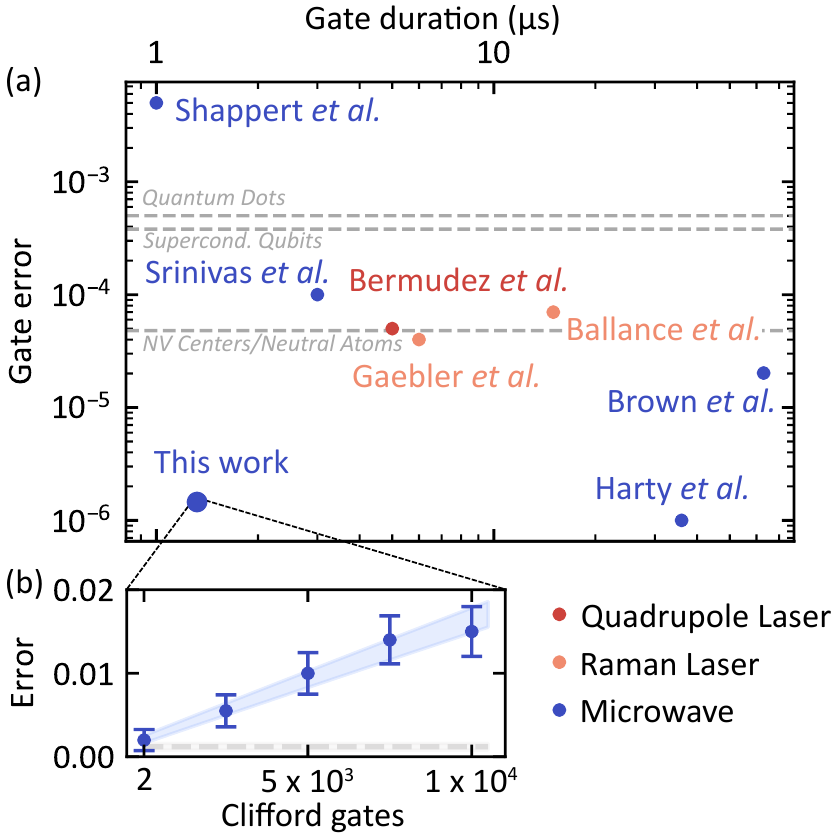}
    \caption{
    \textbf{State of the art for non-addressed single-qubit gates.}
    \textbf{(a)} Selection of single-qubit gate errors and durations across different ion control methods~\cite{Shappert_2013,Srinivas_2021,Bermudez_2017,Ballance_2016,Gaebler_2016,Brown_2011,Harty_2014}.
    Quoted gate durations exclude time delays between pulses ($2~\upmu$s in our case); these are not specified in all references, and should be straightforward to eliminate with appropriate hardware.
    We exclude very fast gate demonstrations ($50~$ps and $19~\upmu$s \cite{Ultrafast,Ospelkaus_2011}), for which low or no fildities are measured.
    Typical single-qubit gate errors in other quantum computing platforms~\cite{Qauntum_dots,Rong_2015,Neutral_atoms_2,Jurcevic_2021} are shown with gray dashed lines.
    \textbf{(b)}
    Randomized benchmarking of single qubit gates in our system.
    Blue dots show the increase in error of a sequence of Clifford gates versus the sequence length.
    The gray dashed line shows the state preparation and measurement (SPAM) error of $1.2(4)\times 10^{-3}$.
    A fit to the data (blue shaded area) yields an average Clifford gate error of $1.5 (1)\times 10^{-6}$.
    }
    \label{fig:comparison}
\end{figure}

Experiments are carried out using a micro-fabricated segmented-electrode surface Paul trap with an on-chip microwave resonator generating a microwave field for the ions trapped at a height of $40~\upmu$m~\cite{trap}.
The trap is operated at room temperature for the single-ion experiments and at ``warm cryogenic'' temperature (100K) for addressing experiments (which improves two-ion trapping lifetime).
Our qubit is defined by the hyperfine levels $\ket{F =4,M =1 }$ and $\ket{F =3,M =1 }$ in the ground state manifold $4\text{S}_{1/2}$ of $^{43}\text{Ca}^+$, which form a clock transition at our static magnetic field strength of 28.8 mT.
Further details, notably concerning state preparation and readout can be found in Ref.~\cite{trap}.
Logical operations are driven by the microwave drive chain described in Supplementary Sec. S1.
Using a different hyperfine transition as a qubit, lowering the ion-height and with resonant enhancement of the microwaves enables our surface trap to perform gates on a single ion on sub-microsecond timescales, whilst maintaining fidelities consistent with the state of the art across all quantum computing platforms.
We illustrate the landscape of single qubit gate fidelities and durations in Fig.~\ref{fig:comparison} (a) with a selection of results across different ion manipulation protocols and quantum computing technologies.

To measure gate errors, we use randomized benchmarking (RB)~\cite{RBM}.
The qubit is subjected to a sequence of pseudorandom Clifford gates which combined perform a known Pauli gate.
Each Clifford is decomposed into $\pi/2$ and $-\pi/2$ pulses in the $\hat{\sigma}_{\text{x}}$ and $\hat{\sigma}_{\text{y}}$ directions with an average of $2.2$ pulses per Clifford.
The probability of measuring the expected state at the end of a sequence decays towards $50~\%$ as the number of applied Clifford gates increases.
Measuring this decay gives us a measure of the average error per Clifford gate.
In this experiment, the average single-qubit Clifford gate error is measured to be $1.5 (1)\times 10^{-6}$, see Fig~\ref{fig:comparison} (b).
The average Clifford gate duration is of $1.32~\upmu$s, arising from an average of 2.2 $\pi/2$-pulses per Clifford gate with a 600ns pulse time, excluding technically imposed $2~\upmu$s inter-pulse delays.
A summary of all known error sources is presented in Table ~\ref{tab:error_budget}.
The dominant contribution to the error budget is the decoherence time, which we measure through memory benchmarking~\cite{IRBM,Sepiol_2019} to be $T_\text{2}^{\ast\ast} = 4.6 (2)~\text{s}$.
Here we introduce the notation $T_2^{\ast \ast}$ to represent the effective decoherence time constant in the {\em small error} regime~\cite{Sepiol_2019}.
The error due to decoherence is increased by the need for a $2~\upmu\text{s}$ delay time after each $0.6~\upmu\text{s}$ $\pi/2$-pulse, a purely technical limitation imposed by the rate at which our field programmable gate array controller (FPGA) can output events to our arbitrary waveform generator (AWG).
The second largest source of error is the thermal occupation of the in-plane secular mode of ion motion.
As the ion moves in-plane parallel to the trap surface, the amplitude of the microwaves changes, and with it the amount of rotation driven on the Bloch sphere.
For slower gates, where the position of the ion performs many oscillations around its equilibrium during a gate, the average Rabi frequency in a gate will remain constant.
%
This effect becomes more significant in our system because the Rabi frequency (520 kHz) approaches the in-plane mode frequency (5.66 MHz).
However, even a worst-case prediction yields a non-limiting $2.4 \times 10^{-7} $ average gate error across a 10,000 Clifford gate sequence.
Methods used to estimate other errors are provided in Supplementary Sec. S3.

\begin{table}[]
\begin{tabular}{l|c}
Error source            & Error ($/ 10^{-6}$ )                 \\ \hline
\rule{0pt}{2.5ex}Decoherence $T_\text{2}^{\ast\ast}$     & 0.42    \\
In-plane motion         & 0.24    \\
microwave/laser leakage        & 0.084    \\
Amplitude stability     & 0.081    \\
Detuning                & 0.075     \\
AC Zeeman shift         & 0.006     \\
Spectator state excitation & 0.003    \\
\hline\hline\rule{0pt}{2.5ex}Simulated error         & 0.91                   \\
Measured error            & 1.5
\end{tabular}
\caption{
\textbf{Single-qubit gate error budget.}
 Errors are simulated for a $\pi/2$-pulse and then scaled by the average number of pulses in a Clifford gate ($2.2$ in our implementation).
 Decoherence during inter-pulse delays ($2~\upmu$s) is also included.
}
\label{tab:error_budget}
\end{table}

\begin{figure}[b]
    \centering
    \includegraphics[width=0.43\textwidth]{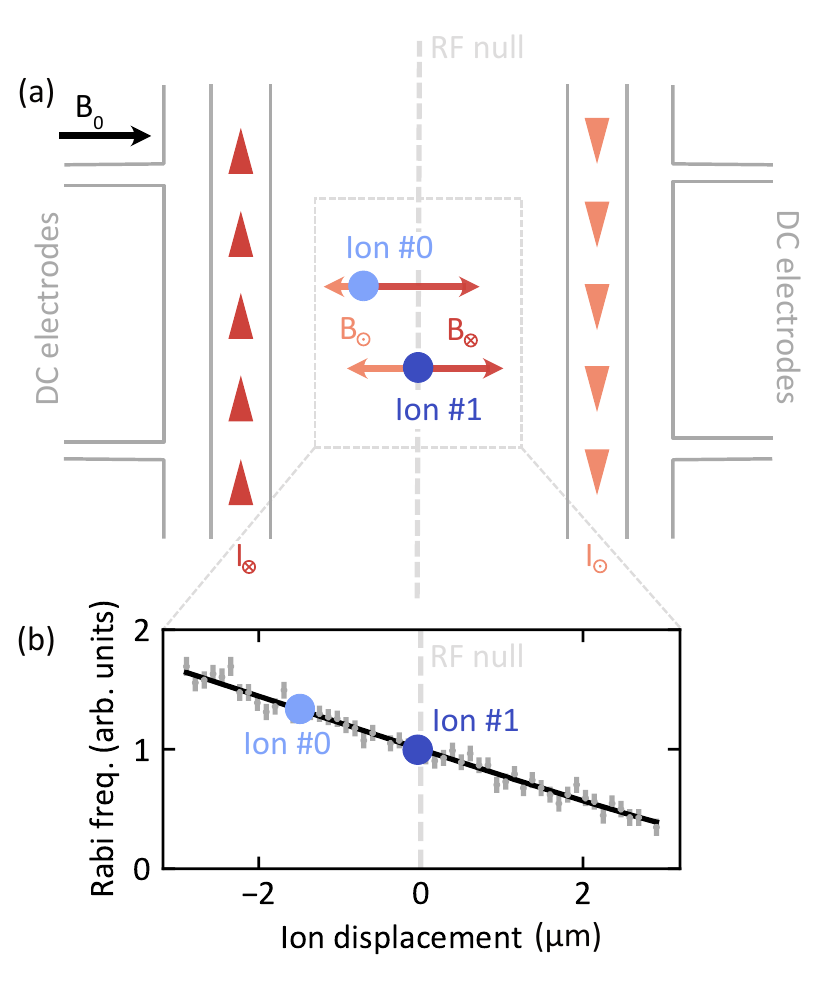}
    \caption{
        \textbf{Surface trap design enabling a microwave field gradient for qubit addressing (not to scale).}
    \textbf{(a)} Schematic top view of the surface trap.
    Two out-of-phase microwave currents (red/orange arrow heads) generate magnetic fields (red/orange arrows) parallel to the quantization axis $\textbf{B}_0$ which drive qubit transitions.
    Destructive interference of the $\pi$-components of these fields leads to a different field for each ion (blue dots) of a twisted crystal.
    \textbf{(b)} Measured qubit Rabi frequencies (gray error bars) for different displacements of an ion from the RF null (gray dashed line).
    In the addressing experiment, we twist a two-ion crystal by $15^\circ$, such that each ion experiences a different Rabi frequency.
    }
    \label{fig:B_vs_x}
\end{figure}

Our fast and high-fidelity single-qubit gates enable the use of a multi-pulse scheme to address single ions.
This scheme relies on the large magnetic field gradient provided by the microwave electrode layout~\cite{trap}.
As shown in Fig.~\ref{fig:B_vs_x}, counter-propagating microwave currents lead to destructively interfering fields along the quantization axis.
This results in a large gradient in the field component required to drive the qubit transition.
For the 130 mW input power used to drive a $\pi/2$-rotation (on par with typical powers used in the addressing pulse scheme) this gradient is $11.7~$T/m.
By changing the voltages of the segmented trap DC electrodes, the trapping potential can be twisted, such that the ions are placed at different locations in the Rabi frequency gradient.
By tuning the amplitudes and phases of a train of microwave pulses (all with identical temporal shape) we can use the differential Rabi frequency to construct an arbitrary pair of different single-qubit gates on the two ions.
Such a pair of gates can be described by the unitary $G_0\otimes G_1$,
\begin{equation}
    G_k=\begin{pmatrix}
        e^{i\delta_k}\cos{\frac{\theta_k}{2}} & e^{i\phi_k}\sin{\frac{\theta_k}{2}} \\
        e^{-i\phi_k}\sin{\frac{\theta_k}{2}} & e^{-i\delta_k}\cos{\frac{\theta_k}{2}}
        \end{pmatrix}\ ,
    \label{eqn:space}
\end{equation}
where $k=0,1$ indexes the ions.
Each unitary $G_k$ has three parameters: $\phi_k$, $\delta_k$ and $\theta_k$, totalling 6 parameters per gate pair.
A single resonant microwave pulse drives this unitary evolution on both ions with a few constraints.
With resonant driving, we have $\delta_0=\delta_1=0$, the phase $\phi$ of the microwaves sets $\phi_0=\phi_1=\phi$, and the relative amount of rotation induced in the qubit states is fixed through $\theta_k = \pi A/A_k^\pi$, determined by the pulse amplitude $A$ relative to the amplitude $A_k^\pi$ required to perform a $\pi$ rotation on ion $k$.
A pulse of amplitude $A$ and phase $\phi$ thus drives the unitary $R_0\otimes R_1$,
\begin{equation}
    R_k=\begin{pmatrix}
        \cos\frac{\pi A}{2A_k^\pi} & e^{i\phi}\sin\frac{\pi A}{2A_k^\pi} \\
        e^{-i\phi}\sin\frac{\pi A}{2A_k^\pi} & \cos\frac{\pi A}{2A_k^\pi}
        \end{pmatrix}\ .
\end{equation}
For each pulse, we therefore have two degrees of freedom to adjust, so at least three pulses are required to match the 6 parameters of the desired pair of gates.
In practice, we use four pulses instead of three such that two additional degrees of freedom permit empirical minimization of the susceptibility to certain errors.
The amplitudes and phases of the pulses are calculated numerically using a least-squares method (see Supplementary Sec. S2).

\begin{figure}
\centering
\includegraphics[width=0.43\textwidth]{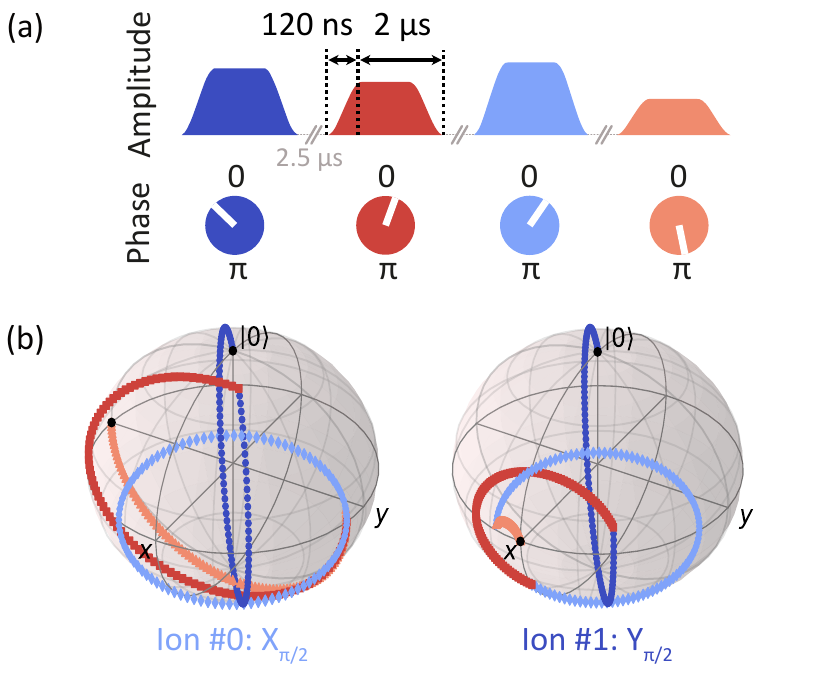}
\caption{
\textbf{Single-ion addressing scheme.}
\textbf{(a)} Decomposition of a pair of addressed gates (X$_\frac{\pi}{2}$ on ion $\#0$ and a Y$_{\frac{\pi}{2}}$ on ion $\#1$) into four microwave pulses.
For each pulse both amplitude and phase is varied, and due to the difference in Rabi frequencies experienced by each ion, they will undergo different amounts of rotation on the Bloch sphere.
\textbf{(b)} The outcome of the pulse sequence is illustrated with both qubits starting in the $\ket{0}$ state.
Ion $\#0$ and ion $\#1$ undergo trajectories on the Bloch sphere ending in the $\ket{+}$ and the $\ket{+\text{i}}$ state, corresponding to X$_\frac{\pi}{2}$ and Y$_\frac{\pi}{2}$ gates respectively.
}
\label{fig:scheme}
\end{figure}

We illustrate the scheme in Fig.~\ref{fig:scheme} (a), where the implementation of a X$_\frac{\pi}{2}$ gate on ion $\#0$ and a Y$_\frac{\pi}{2}$ gate on ion $\#1$ is shown.
The corresponding trajectories on the Bloch spheres shown in Fig.~\ref{fig:scheme} (b) demonstrate that despite the axis of rotation being the same for both qubits in each pulse, the differential Rabi frequency $\left(\Omega_1/\Omega_0 = 0.80 \right)$ is ultimately sufficient to reach the target state.
This scheme is implemented with $2.12~\upmu \text{s}$ pulses of varying amplitude and phase, resulting in an addressed gate duration of $8.48~\upmu$s (excluding inter-pulse delays).
Each pulse is ramped on and off with a $\sin^2\left(t\pi/2t_R\right)$ shape (with $t_R=120~\text{ns}$) to avoid exciting spectator hyperfine transitions.

\begin{figure}
    \centering
    \includegraphics[width=0.43\textwidth]{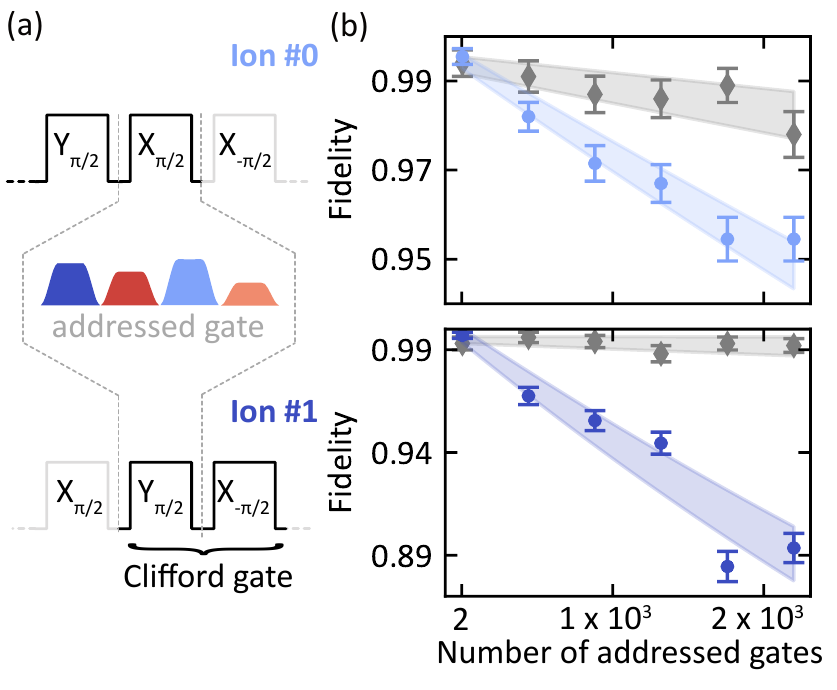}
    \caption{
    \textbf{Simultaneous randomized benchmarking of addressed gate pairs.}
    \textbf{(a)} Clifford gates used in the RB sequence are decomposed into combinations of
    X$_{\pm\frac{\pi}{2}}$, Y$_{\pm\frac{\pi}{2}}$ and identity gates I.
    Each pair of X$_{\pm\frac{\pi}{2}}$, Y$_{\pm\frac{\pi}{2}}$ or I gates is decomposed into four physical pulses using the addressing scheme.
    \textbf{(b)} The blue data and fit (dots and shaded area) show the measured decrease in fidelity with the number of X$_{\pm\frac{\pi}{2}}$, Y$_{\pm\frac{\pi}{2}}$ and I gates in a RB sequence.
    The gray data and fit (dots and shaded area) correspond to the same protocol, but where all physical pulses are replaced with time delays of the same duration; the latter constitutes a measure of SPAM error, which is subtracted from the former measurement to obtain the gate error.
    These measurements give an average addressed gate error of $1.6 (3)\times 10^{-5} $ for ion $\#0$ and $5.2 (7) \times 10^{-5} $  for ion $\#1$.
    }
    \label{fig:addressing_rbm}
\end{figure}

To determine the quality of the addressing scheme, we perform RB on both ions simultaneously, each ion being subject to an independent sequence of Clifford gates.
As with RB on a single qubit, Clifford gates are decomposed into X$_{\pm\frac{\pi}{2}}$ and Y$_{\pm\frac{\pi}{2}}$ gates.
A pair of X$_{\pm\frac{\pi}{2}}$ or Y$_{\pm\frac{\pi}{2}}$ gates (one gate applied to each ion simultaneously) is finally decomposed into four physical pulses using the addressing scheme.
This is illustrated in Fig.~\ref{fig:addressing_rbm} (a).
Whilst the number of Clifford gates applied to each ion is the same in a single shot of the experiment, the number of underlying X$_{\pm\frac{\pi}{2}}$ and Y$_{\pm\frac{\pi}{2}}$ gates necessary to implement all the Clifford gates may differ.
The shorter sequence is padded with identity gates I to account for this.
These identities are implemented with the same composite pulse method as the $\pi/2$-gates.
Even though there are in principle 25 different pairs of X$_{\pm\frac{\pi}{2}}$, Y$_{\pm\frac{\pi}{2}}$ and I gates that a Clifford gate decomposition could require, we make use of the global phase in the microwave pulse sequence to reduce the number of pulse sequences that we need to compute and calibrate.
For example, subjecting the ions to a pulse sequence implementing gates X$_\frac{\pi}{2}$ on ion $\#1$ and Y$_{\frac{\pi}{2}}$ on ion $\#1$, but with the microwave phase shifted by $45^\circ$, realizes a Y$_\frac{\pi}{2}$ on ion $\#0$ and a X$_{-\frac{\pi}{2}}$ on ion $\#1$.
This reduces the number of pulse sequences required for RB to only six.
For each required sequence, we make use of the fourth pulse to test multiple sequences and to select the one which offers the best fidelity (see Supplementary Sec. S2).

Whilst the pulse sequences only implement X$_{\pm\frac{\pi}{2}}$, Y$_{\pm\frac{\pi}{2}}$ or I gates in this experiment, they can in principle implement arbitrary gates.
Hence, the figure of merit for this scheme is the fidelity of the addressed operation resulting from the sequence of four pulses.
We therefore measure RB sequence lengths in terms of the number of X$_{\pm\frac{\pi}{2}}$, Y$_{\pm\frac{\pi}{2}}$ and I gates, and quote fidelities for these addressed operations, rather than for the Clifford gates that are composed of several such addressed gates.
In Fig.~\ref{fig:addressing_rbm} (b), we show the evolution of the sequence fidelity as a function of the number of gates.
The states of the ions are read out individually using ion shuttling~\cite{MariusThesis}.
The resulting errors per addressed gate are $1.6 (3)\times 10^{-5} $ and $5.2 (7) \times 10^{-5} $ for ion $\#0$ and ion $\#1$ respectively, i.e. an average addressed gate error of $3.4 (3)\times 10^{-5}$.
In Table~\ref{tab:Addressing}, we compare this error, as well as the gate duration, to previous microwave addressing experiments.

\begin{table}[t!]
    \begin{tabular}{r|c|c|c|c}
               & Error&Crosstalk&Duration& $\#$ ions          \\
               & ($/ 10^{-3} $) &($/ 10^{-3} $)&($\upmu$s)&              \\ \hline
    \rule{0pt}{2.5ex}This work     & \multicolumn{2}{c|}{0.03}& 8.5 &2   \\ \cline{2-3}
    \rule{0pt}{2.5ex}Craik et al. \cite{Craik_2017}     & -  & 3 & 50--90&2 \\
    Randell et al. \cite{Randell_2015}         & - & 5 & 550&2\\
    Piltz et al. \cite{Piltz_2014}    & $\ge$5   &0.03--0.08&25&8   \\
    Piltz et al. \cite{Piltz_2014}     & $\ge$5   &0.06--0.23&9&8  \\
    Warring et al. \cite{Warring_2013}     &  -   & 0.6--1.5&  50&2 \\

    \end{tabular}
    \caption{
\textbf{Trapped-ion microwave addressing state of the art.}
    Comparison of single-qubit addressed gates for error, nearest-neighbour crosstalk and gate duration across different microwave-driven ion trap experiments~\cite{Piltz_2014,Craik_2017,Warring_2013,Randell_2015}.
    Quoted gate durations exclude time delays between pulses.
    Where crosstalk error was not measured through RB, crosstalk for a $\pi$-pulse is quoted.
    Our simultaneous benchmarking approach does not distinguish between gate error and crosstalk as gates are executed in parallel.
    }
    \label{tab:Addressing}
\end{table}

The measured error is larger than expected from scaling the single-qubit gate error: by linearly scaling the single-qubit gate error with pulse duration, we would expect an error per gate in the addressing scheme of $9 \times 10^{-6}$.
The excess error can however be explained by drift in the microwave amplitude.
We monitor the microwave amplitude by measuring the average state of one of the ions in the twisted crystal after being subjected to $101$ $\pi/2$-pulses.
The drift measured over tens of minutes is sufficient to limit the gate fidelity (see Supplementary Sec. S4).
We find that this level of drift in the microwave field amplitude is significantly worse than that measured for a single ion (Sec. S3 C), and we suspect that the drifts may be associated with position drifts of the ions which are present for two ions in the twisted configuration.
This could be due to the larger, and asymmetric, DC voltages which are required to twist the ion crystal, leading to greater susceptibility to common-mode voltage noise.
In conclusion, we have demonstrated fast ($1.32~\upmu$s) and high-fidelity ($1.5 \times 10^{-6} $ error) single-qubit gates driven by microwave near-field radiation.
This level of performance has enabled a high-fidelity single-ion addressing scheme using optimized pulse sequences, in which gates can be carried out simultaneously on two ions within the same potential well with an average error of $3.4 \times 10^{-5}$.
This surpasses the best performance achieved by -- technically much more demanding -- optical addressing approaches~\cite{Crain_2014,binaimotlagh2023} and appears to be the lowest error reported across all physical platforms for in-register single-qubit addressing.
With this work we demonstrate that nulling the effects of microwave fields -- either by frequency selection~\cite{Piltz_2014,Warring_2013,Randell_2015}, field cancellation~\cite{Craik_2017,Warring_2013}, or sideband manipulations~\cite{Warring_2013,sutherland2022,srinivas2022} -- is not a strict requirement for addressing individual qubits.
The addressing scheme could theoretically be extended to more than two ions, with the number of required pulses scaling linearly with the number of qubits.
The parallel nature of the scheme means that only $3N/2$ pulses are required to address all ions of an $N$-ion crystal.
Each ion adds three degrees of freedom in the target unitary, and each pulse offers two control parameters (amplitude and phase), hence the factor $3/2$.
Our scheme therefore scales as a sequential addressing schemes (e.g. single laser addressing or microwave field nulling) and is potentially faster than microwave~\cite{Warring_2013,Craik_2017} or laser~\cite{Quantinuum} addressing methods requiring slower shuttling operations.
Since this addressing method does not induce a difference in the microwave amplitude gradient experienced by each ion, single and two-qubit gates~\cite{harty2016} can be interleaved without changing the ion positions.
In the short term, this could enable the implementation of RB for two-qubit gates without the limitation of using subspace benchmarking methods~\cite{Baldwin_2020}.
The implementation of the scheme could be further simplified by ``embedding'' the differential Rabi frequency in the surface trap design, e.g. by angling the microwave delivery electrodes (which could be ``buried'' beneath the trapping electrodes in a multi-layer design~\cite{Hahn_2019}) with respect to the RF electrodes.
This would also avoid inducing RF micromotion when twisting the ion crystal.
Finally, the scheme could be used on other ground state transitions to selectively move the population out of the computational basis of one of the ions and then into shelf states (through the $\text{S}_{1/2}-\text{D}_{5/2}$ quadrupole transition), enabling individual ion readout without the need for ion shuttling or tightly focused laser beams.
More generally, the efficient composite pulse scheme which we have introduced could be used for individual qubit addressing in any physical system where a modest differential Rabi frequency between qubits can be engineered, for example to allow multiple qubits to share a single microwave control line.

\vspace{8mm}
\textbf{Acknowledgments:}
This work was supported by the U.S. Army Research Office (ref. W911NF-18-1-0340) and the U.K. EPSRC Quantum Computing and Simulation Hub.
M.F.G. acknowledges support from the Netherlands Organization for Scientific Research (NWO) through a Rubicon Grant.
A.D.L. acknowledges support from Oxford Ionics Ltd.

%% file: SI.tex

\section{Experimental Setup}
\label{sec:experimental_setup}

\begin{figure}[h!]
\centering
\includegraphics[width=0.86\textwidth]{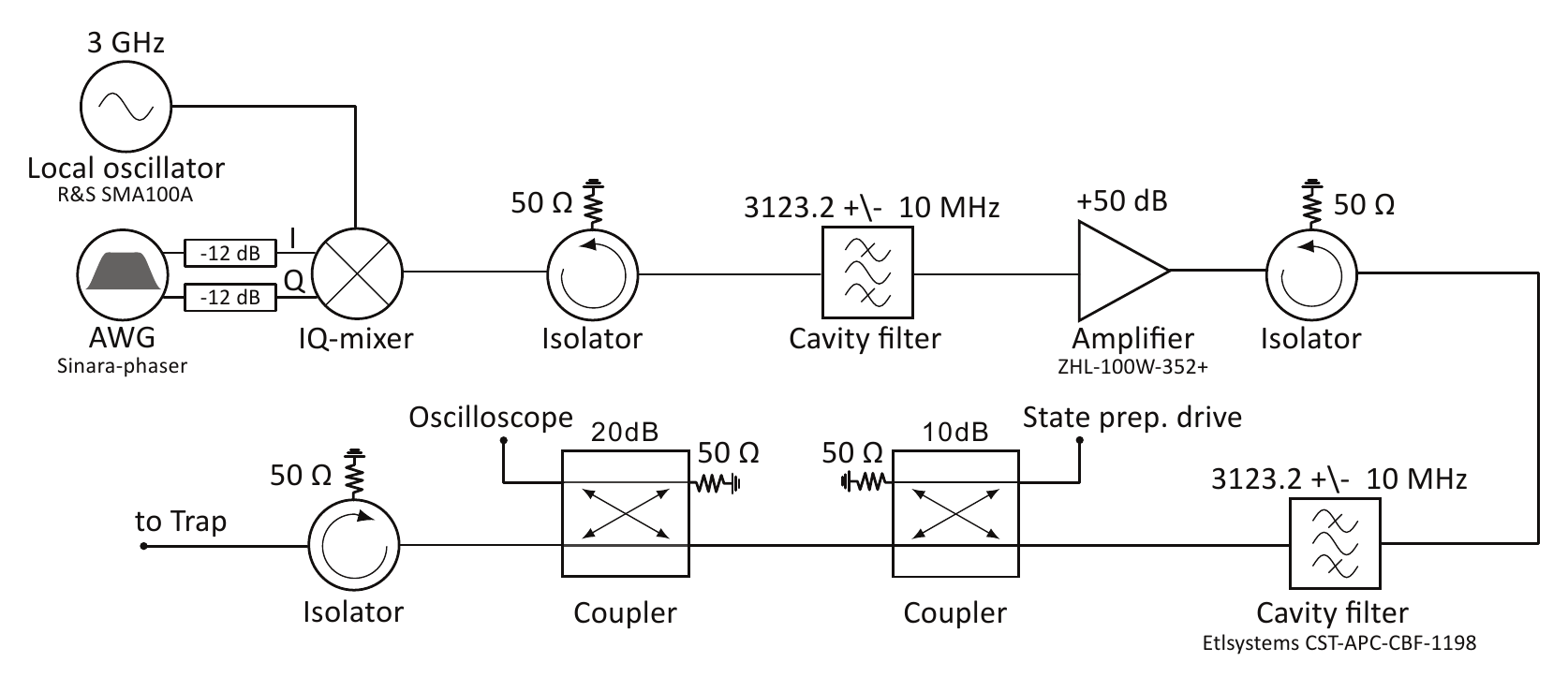}
\caption{
\textbf{Microwave setup}
}
\label{fig:experimental_setup}
\end{figure}

The pulses used to drive single-qubit gates are generated by an arbitrary wave form generator (AWG, Sinara-Phaser) which is controlled by an FPGA (Sinara-Kasli) programmed amongst other devices through the ARTIQ framework~\cite{artiq}.
These pulses are up-converted to microwave frequencies and subsequently filtered and amplified as shown in Fig.~\ref{fig:experimental_setup} before being combined with a separate microwave drive chain providing state preparation and readout pulses.
These microwave signals are then guided towards the surface ion trap for which further detail is provided in Ref.~\cite{trap}.

\FloatBarrier
\section{Calibration Procedure}
\label{sec:Calibration_Procedure}
To calculate the amplitude and phases of the pulses used in the addressing gates, we first determine the difference in Rabi frequency between the two ions.
After twisting the two-ion crystal by $15^\circ$, the microwave amplitude $\text{A}_{k}^\pi$ required to produce a spin flip on ion $k$ is calibrated for both ions.
We then use a least-square method to minimize a cost-function measuring gate error.
Calculations are done with rotation matrices in the special orthogonal group $\text{SO}(3)$, where a gate is represented as a rotation on the Bloch sphere.
The cost function is given by
\begin{equation}
    \sum_{k=0}^1\Bigg\|\left(\prod_{j=1}^{4}\textbf{R}_{\phi_j}\left(\dfrac{\text{A}_j}{\text{A}_{k}^\pi}\pi\right)\right) - \textbf{G}_k\Bigg\|_\mathrm{HS}
    \label{eqn:costfunction}
\end{equation}
where $\|\cdot\|_\mathrm{HS} $ denotes the Hilbert-Schmidt norm, $\textbf{R}_{\phi_j}(\theta_j)\in\text{SO}(3)$ encodes the rotation on the Bloch sphere performed by pulse $j$ (through an angle $\theta_j$ with respect to an axis on the equator defined by the angle $\phi_j$), and matrix $\textbf{G}_k\in\text{SO}(3)$ is the rotation matrix corresponding to the desired gate on the ion $k$.
Only pulse amplitudes $\text{A}_j$ and phases $\phi_j$ are varied in the minimization routine.

\begin{figure}[h!]
\centering
\includegraphics[width=0.86\textwidth]{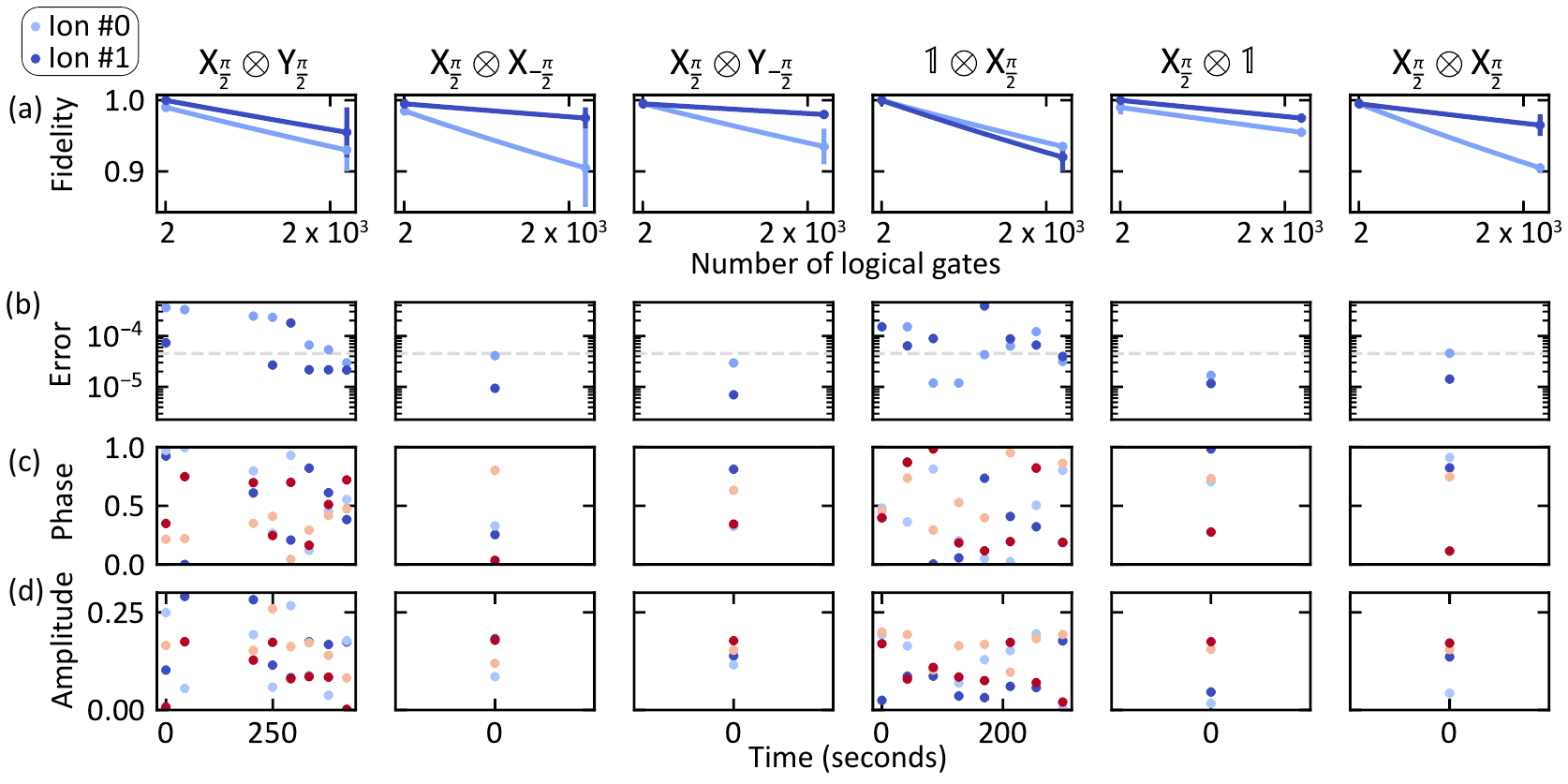}
\caption{
\textbf{Calibration procedure for single-ion addressing.}
\textbf{(a)} RB results for the pulse sequence which produced a sub-threshold error and triggered the calibration of the next pair of gates.
\textbf{(b)} RB error for each tested pulse sequence in the calibration.
The gray dashed line corresponds to the threshold which triggers the calibration of the next pair of gates or the end of the procedure.
\textbf{(c)} Phase of each pulse in the generated sequences (in units of $2 \pi$).
The colors in this pane and the following match the coloring shown in figure \ref{fig:scheme}.
\textbf{(d)} Amplitude (a.u.) of the pulses of the generated sequence.
}
\label{fig:calibration}
\end{figure}

Since we use an additional pulse to the three strictly required to construct a pair of gates, different initial guesses in the optimization procedure can produce different pulse sequences.
Comparing different pulse sequences experimentally (for the same target gate) reveals that the error can vary over an order of magnitude across different pulse sequences.
We suspect that different pulse sequences amplify or attenuate coherent error sources.
We exploit this fact in our calibration routine by generating different pulse sequences and testing them experimentally to maximize fidelity.
Specifically, for a given pair of target gates, we generate a pulse sequence and measure its fidelity by running the same RB sequences on both ions simultaneously.
If the error (averaged over two ions) exceeds a threshold of $5\times10^{-5}$, we repeat the process with another generated gate sequence, otherwise, the sequence is memorized and the process is applied to the next target gate pair.
Once the pulse sequence for all six required gate pairs are determined, we carry out the RB experiment as described in the main text.
The calibration procedure for the main text RB data is shown in Fig.~\ref{fig:calibration}.

\section{Single qubit gate calibration and errors}
\label{sec:errors}
In this section, we provide further details on the construction of the single-qubit error budget table (Table~\ref{tab:error_budget}).

\subsection{Qubit detuning}
Because of the magnetic field component of the radio-frequency trapping field, the effective energy splitting of our qubit is shifted through the AC Zeeman effect.
We calibrate this shift by measuring the detuning at which maximum population transfer is achieved for a weak, 2 ms long, microwave pulse.
The length of the pulse provides good frequency resolution, and its weakness ensures negligible AC Zeeman shift from the pulse itself will be present.
To verify this calibration, we measure the Clifford gate error with RB as the detuning is swept as shown in Fig.~\ref{fig:rbm_sweep} (a).
This demonstrates that our set-point (the origin in the x-axis) is close to optimal.
Without this calibration, the detuning induced by the RF trapping field (120 Hz) would dominate the gate error.
The detuning error in Table 1 is calculated from the typical day-to-day drift of this calibrated quantity.

\subsection{Spectator state excitation}
Given the peak amplitude of the Rabi frequency ($\sim0.5$ MHz), relative to the frequency difference between the qubit transition and transitions to spectator states ($\sim100$ MHz), a significant amount of the state population $\left(\sim10^{-4}\right)$ resides in spectator states during a gate.
By ramping up our pulses much slower than the period of a Rabi oscillation with the spectator states, we adiabatically transition in and out of a dressed state basis spanning mostly the qubit and its four neighboring spectator states.
More specifically, the pulse length (600 ns) includes two  $t_R=120~\text{ns}$ on/off ramping periods with a $\sin^2\left(\pi t / 2 t_R\right)$ shape.

\subsection{Pulse amplitude}
\label{sec:single_microwave_stability}
With a fixed temporal shape, the remaining parameter used to calibrate the pulse to a  $\pi/2$-rotation is the microwave amplitude.
We calibrate the microwave amplitude by maximizing the probability of returning to a prepared state after up to 1024 repetitions of the pulse.
To verify this calibration procedure, we measure the gate error with RB as a function of deviation from the calibrated parameter as shown in Fig.~\ref{fig:rbm_sweep} (b).
The drift in microwave amplitude is measured by monitoring the qubit state after performing $1049$ $\pi/2$-pulses.
The result, shown in Fig.~\ref{fig:ampl_stability}, shows that the impact of drift is orders of magnitude lower than single qubit gate error.
Notably, loading events do not have a significant effect on the Rabi frequency.

\subsection{Microwave AC Zeeman shift}
The dressing of the qubit by spectator states also affects the qubit frequency during the gates through the AC Zeeman effect.
As we are unable to ramp the microwave frequency with the microwave amplitude, we rather compensate the shift with a fixed detuning of the microwaves over the whole duration of the pulse.
In the rotating frame of the microwaves, we then track the phase of the qubit during the inter-pulse delay, where the qubit is unaffected by the Zeeman shift, and account for the resulting phase shift in the next pulse.
The Zeeman shift is calibrated by performing 800 $\pi/2$ pulses with alternating positive and negative phases.
Only if the Zeeman shift is correctly calibrated, these pulses should return the qubit to its initial state, allowing us to determine the optimal detuning of $\sim270$ Hz.
This value is close to the theoretical value for the AC Zeeman shift (283 Hz), which assumes that the $\sigma_+$ and $\sigma_-$ polarized components of the microwave field are equal.
The error resulting from typical day-to-day drift of this parameter is negligible and quoted in Table 1.
We verify this calibration technique by detuning the microwaves applied during the pulse (with respect to the calibrated optimum), and monitoring the impact on RB error, see Fig.~\ref{fig:rbm_sweep} (c).

\subsection{Decoherence}
\begin{figure}[h!]
\centering
\includegraphics[width=0.86\textwidth]{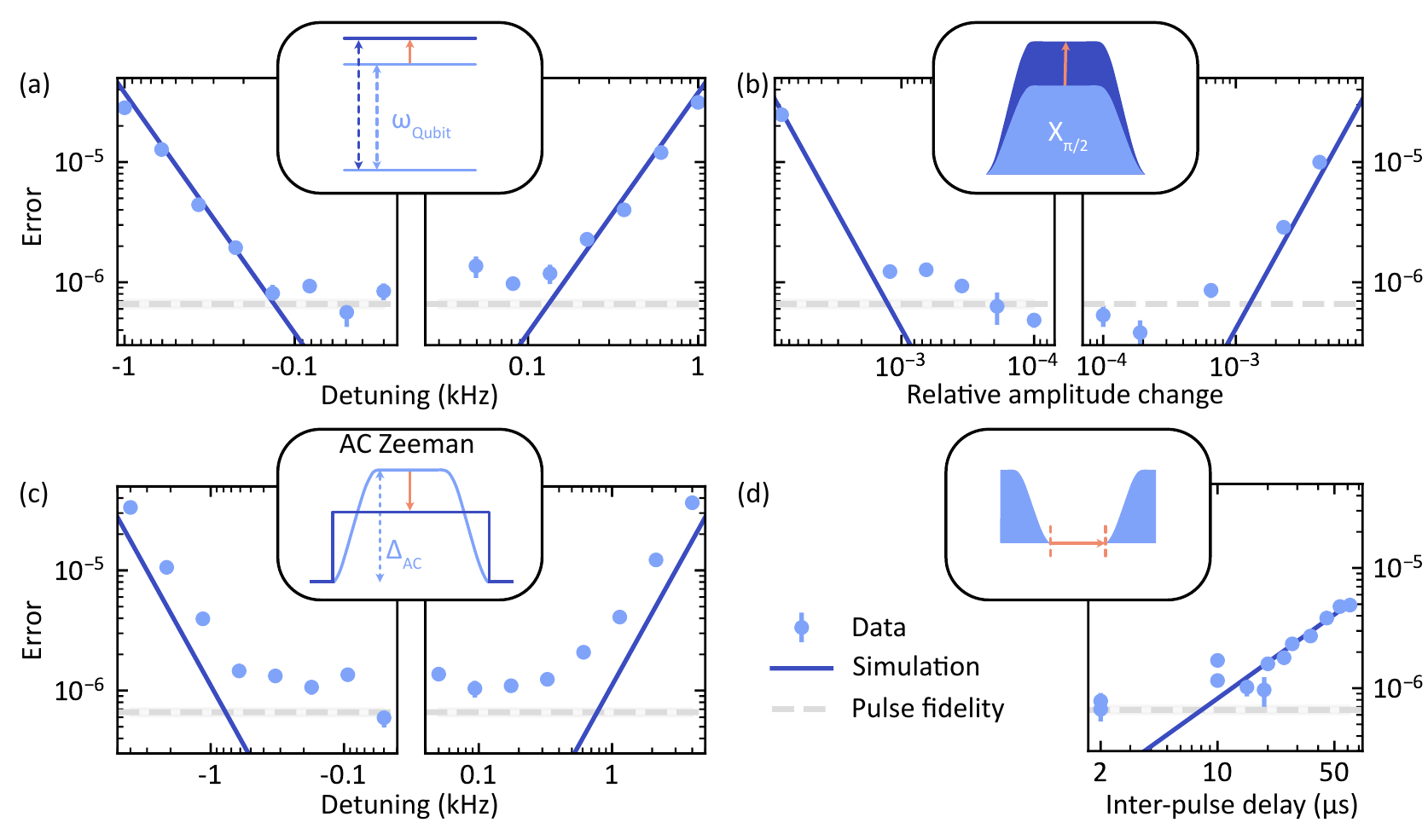}
\caption{
\textbf{Single-qubit gate error model verification.}
Points correspond to the average $\pi/2$ rotation error extrapolated from Clifford gate infidelities measured with RB.
The blue solid line is a theoretical estimate of the average $0.6~\upmu$s pulse error with $2~\upmu$s inter-pulse delay and the gray dashed line corresponds to the pulse error measured in Fig.1(b).
In each panel, the relation between error and a different gate parameter is measured and simulated.
These parameters are:
\textbf{(a)} offset from the calibrated qubit frequency,
\textbf{(b)} offset in calibrated microwave amplitude,
\textbf{(c)} offset in detuning during the pulse (which attempts to compensate for the AC-Zeeman shift),
\textbf{(d)} inter-pulse delay.
}
\label{fig:rbm_sweep}
\end{figure}

To estimate the impact of decoherence on the small time-scales which matter to an RB experiment, we vary the inter-pulse delay and measure its impact on gate error, see Fig.~\ref{fig:rbm_sweep} (d).
Assuming that the origin of the resulting error is dephasing, which introduces an error linear in inter-pulse delay, we get an estimate of the decoherence time $\text{T}_2^{\ast\ast} = 4.6 (2)~\text{s}$.
The minimum inter-pulse delay required by our control system, as well as the dephasing occurring during the pulse itself, is the largest contribution to our gate error that we were able to identify.

\subsection{Microwave and laser leakage}
Leakage of radiation affecting the qubit state is measured by preparing the qubit in either logical state, waiting up to 1 second, and measuring the change in qubit population.
We measure a loss of state population of $\sim 2$ \% over the course of a second, corresponding to the Clifford gate error quoted in Table 1.

\subsection{Thermal motion}
In the worst case scenario of a 10,000 Clifford gate sequence, anomalous heating of 400 quanta of motion per second lead to an average thermal occupation of the in-plane radial mode of 23 quanta by the end of the sequence.
Given the effective Lamb-Dicke parameter of $\eta=8\times 10^{-4}$~\cite{trap}, and simulating a $\pi/2-$pulse in the presence of ion-motion coupling, we extrapolate from small Hilbert space and low thermal occupation simulations to obtain an error of $2.4 \times 10^{-7} $ per Clifford gate over the entire sequence.

\begin{figure}[h!]
\centering
\includegraphics[width=0.86\textwidth]{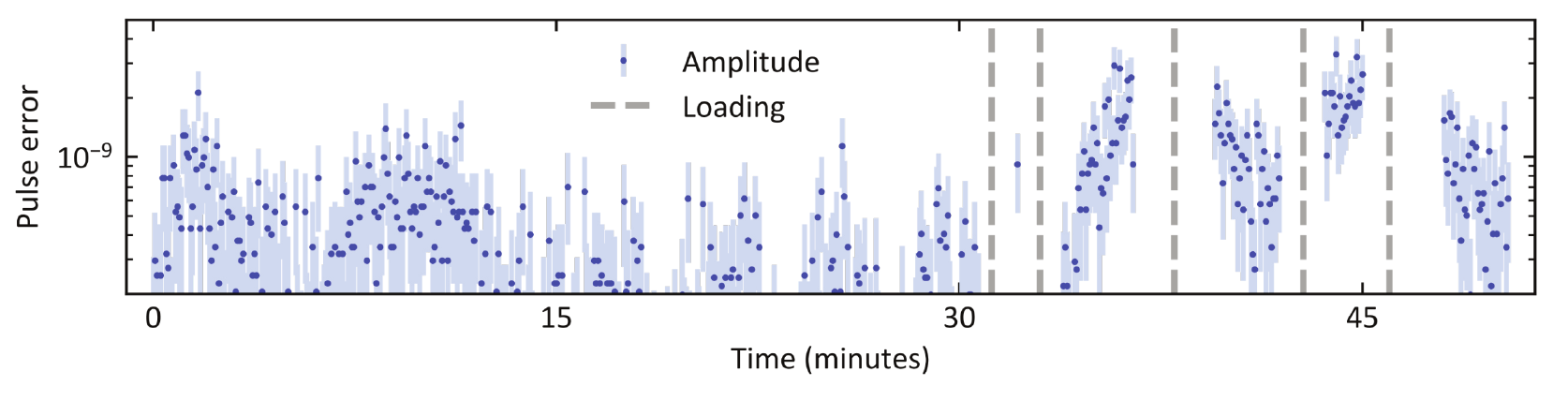}
\caption{
\textbf{Microwave stability (single-qubit gates).}
By preparing the qubit in state $\ket{0}$ and performing $1049$ $\pi/2$-pulses, drifts in the average final state occupation inform us about drifts in microwave amplitude.
In this graph, the result is converted to the expected pulse error.
Gray dashed lines indicate ion loading events.
}
\label{fig:ampl_stability}
\end{figure}

\FloatBarrier
\section{Microwave stability in the twisted two ion crystal}
\label{sec:MW_stability_addressing}
\begin{figure}[h!]
\centering
\includegraphics[width=0.86\textwidth]{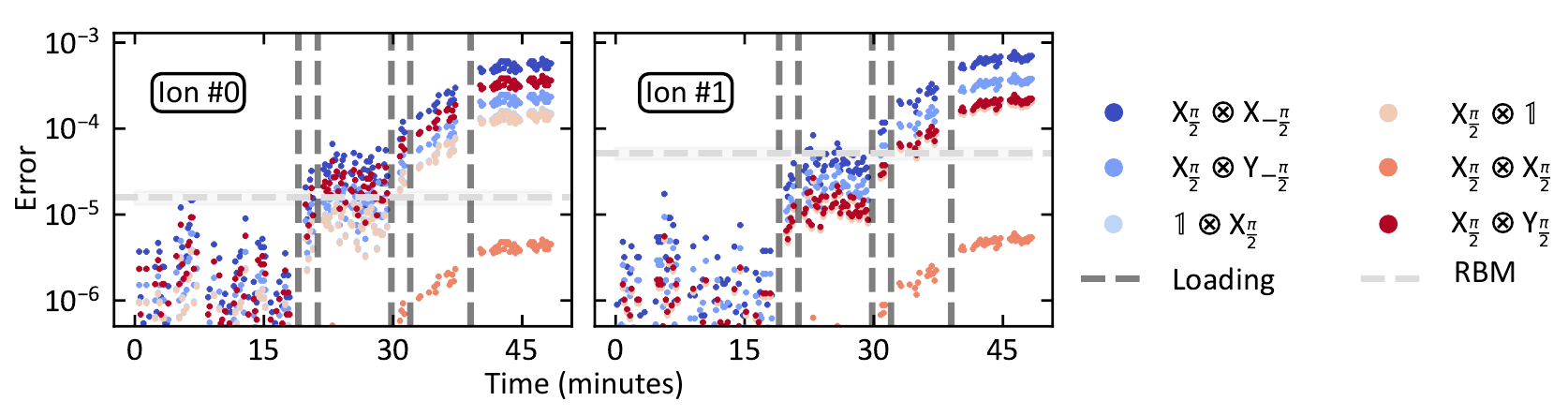}
\caption{
\textbf{Microwave stability (addressed gates).}
Evolution of the different addressing gate errors inferred from the measured drift in microwave amplitude.
The change in microwave amplitude is determined by monitoring the average qubit state of each ion after performing $101$ $\pi/2$ pulses on one of the two ions.
Vertical dashed lines indicate a loading event and horizontal dashed lines corresponds to the logical gate fidelity measured in Fig.~\ref{fig:addressing_rbm}.
}
\label{fig:addressing_stability}
\end{figure}

Contrary to the single ion case, we find that with two ions in a twisted crystal, the microwave amplitude undergoes significant drift on the time-scale of tens of minutes.
As with a single ion, we amplify this effect by monitoring the average qubit occupation after a sequence of $100$ microwave pulses, which perform a $223~\pi/2$ rotation on ion 0 and a $177~\pi/2$ rotation on ion 1.
In Fig.~\ref{fig:addressing_stability}, the drift is converted into an error for the different pulses used in the simultaneous RB shown in Fig.~\ref{fig:addressing_rbm}.
We know from the single ion measurement (see Fig.~\ref{fig:ampl_stability}) that the amplitude of the microwave field at a given position in space is much more stable than the drift shown here.
Therefore, we conclude that the drift lies in a change of position of the ions, probably due to changes in the trapping fields.

